\documentclass[onecolumn,preprintnumbers,amsmath,amssymb]{revtex4}
\usepackage{amsmath, amsthm, amssymb,graphicx,mathtools,algpseudocode,algorithmicx}
\begin{document}
\title{The breakdown of the reaction-diffusion master equation with non-elementary rates}
\author{Stephen Smith}
\affiliation{School of Biological Sciences, University of Edinburgh, Mayfield Road, Edinburgh EH9 3JR, Scotland, UK}

\author{Ramon Grima}
\affiliation{School of Biological Sciences, University of Edinburgh, Mayfield Road, Edinburgh EH9 3JR, Scotland, UK}
\begin{abstract}
The chemical master equation (CME) is the exact mathematical formulation of chemical reactions occurring in a dilute and well-mixed volume. The reaction-diffusion master equation (RDME) is a stochastic description of reaction-diffusion processes on a spatial lattice, assuming well-mixing only on the length scale of the lattice. It is clear that, for the sake of consistency, the solution of the RDME of a chemical system should converge to the solution of the CME of the same system in the limit of fast diffusion: indeed, this has been tacitly assumed in most literature concerning the RDME. We show that, in the limit of fast diffusion, the RDME indeed converges to \emph{a} master equation, but not necessarily the CME. We introduce a class of propensity functions, such that if the RDME has propensities exclusively of this class then the RDME converges to the CME of the same system; while if the RDME has propensities not in this class then convergence is not guaranteed. These are revealed to be elementary and non-elementary propensities respectively. We also show that independent of the type of propensity, the RDME converges to the CME in the simultaneous limit of fast diffusion and large volumes. We illustrate our results with some simple example systems, and argue that the RDME cannot be an accurate description of systems with non-elementary rates. 
\end{abstract}
\maketitle
\noindent

\section{Introduction}
The chemical master equation (CME) describes the fluctuations of molecule numbers in reactive chemical systems which are dilute and well-mixed. In particular, it assumes that the probability of two particles reacting with each other is independent of their relative positions in space, which is strictly true only if diffusion rates are infinitely large, i.e., well-mixed conditions. The CME has in fact been derived from a microscopic physical description under these conditions \cite{gillespie1992rigorous}. 

The applicability of the CME to understand intracellular processes is limited because experiments show that diffusion coefficients inside cells are typically considerably smaller than {\emph{in vitro}} due to macromolecular crowding and other effects \cite{Ellis2001}. The reaction-diffusion master equation (RDME) generalises the CME, in an approximate way, to include diffusion. Space is partitioned into small volume elements (``voxels") of equal volume, each considered to be well-mixed (although globally the space is not well-mixed). In each voxel, chemical reactions occur and diffusion of chemicals occurs as a hopping of particles between neighbouring voxels. The RDME is the Markovian description of this lattice-based reaction-diffusion system. Unlike the CME, the RDME currently has no rigorous microscopic physical basis, but can be justified intuitively, and has been shown to be accurate under certain conditions \cite{smith2015approximate,erban2009stochastic,isaacson2009reaction,gillespie2012simple}.

The RDME and the CME describe the same systems, but the RDME claims greater accuracy in the sense that it contains all the information of the CME while incorporating local spatial effects which are beyond its scope. Since we know the CME to be valid if diffusion rates are fast \cite{gillespie1992rigorous}, a useful test of the validity of the RDME is that it should converge to the CME in the limit of infinite diffusion rates. This convergence seems intuitively likely: if diffusion rates are fast then particles hop between voxels much more frequently than they react, and so the well-mixedness assumption of the CME should be recovered. 

In this paper, we prove that the RDME converges to \emph{a} master equation in the limit of fast diffusion, but remarkably, under certain conditions, this master equation is not the CME. The accuracy of the RDME under these conditions is therefore called into question. The paper is organised in the following way. In Section \ref{Proof} we derive an expression for the master equation to which the RDME converges in the limit of fast diffusion. We subsequently define a class of reaction types (and their corresponding propensity functions) which we call \emph{convergent propensity functions} with the following property: \emph{if a chemical system has exclusively convergent propensity functions, then the RDME will converge to the CME of the same system in the limit of fast diffusion}. If a chemical system has any non-convergent propensity functions, then it will almost surely not converge to the correct CME. In Section \ref{Converg} we show that elementary reactions (including zero, first and second-order reactions) are in the convergent class, while more complex reactions (including Michaelis-Menten and Hill-type) are of the non-convergent class. In Section \ref{Examples} we illustrate our results by applying them to two simple systems: one convergent and one non-convergent. We conclude with a summary and discussion in Section \ref{Discussion}.

\section{Proof}\label{Proof}
In this section we introduce the CME and the RDME and prove that the latter may or may not converge to the former in the limit of fast diffusion. We further prove that independent of the type of propensity, the RDME still converges to the CME in the limit of fast diffusion and of large volumes, taken simultaneously.

\subsection{The CME}

Consider a well-mixed compartment of volume $\Omega$ in which there is a chemical system consisting of $N$ chemical species, $X_1,...,X_N$ involved in $R$ possible chemical reactions where the $j^\text{th}$ reaction has the form:
\begin{equation}\label{CMEreact}
s_{1j}X_1+... + s_{Nj}X_N \xrightarrow{} r_{1j}X_1+...+r_{Nj}X_N,
\end{equation}
where $r_{ij}$ and $s_{ij}$ are the stoichiometric coefficients. The stochastic dynamics of such a system is Markovian and can be described by the CME:
\begin{equation}\label{CME}
\frac{d}{dt}P(\vec{n},t)=\sum_{j=1}^R  \left(\prod_{i=1}^N E_i^{s_{ij}-r_{ij}}-1\right)\hat{a}_j(\vec{n},\Omega)P(\vec{n},t),
\end{equation}
where $\vec{n}=\left( n_1,...,n_N \right)^T$ is the vector of molecule numbers of species $X_1,...,X_N$ respectively, $P(\vec{n},t)$ is the probability that the system is in the state $\vec{n}$ at time $t$, $E_i^x$ is the operator which replaces $n_i$ with $n_i+x$, and $\hat{a}_j(\vec{n},\Omega)$ is the \emph{propensity function} of reaction $j$. The propensity function is formally defined as follows: given that the system is in state $\vec{n}$, then $\hat{a}_j(\vec{n},\Omega) dt$ is the probability that a reaction of index $j$ occurs somewhere in the volume $\Omega$ in the next infinitesimal time interval $[t,t+dt)$ \cite{gillespiereview2007}.

\subsection{The RDME}

Consider the same chemical system as defined by \eqref{CMEreact}, but now taking place in a volume of space $\Omega$ which is divided into $M$ voxels, each of volume $\frac{\Omega}{M}$. The $j^\text{th}$ chemical reaction in voxel $k$ will have the form:
\begin{equation}\label{RDMEreact}
s_{1j}X_1^k+...s_{Nj}X_N^k \xrightarrow{} r_{1j}X_1^k+...+r_{Nj}X_N^k,
\end{equation}
where $X_i^k$ refers to species $X_i$ locally in voxel $k$. Note that because the stoichiometric coefficients $s_{ij}$ and $r_{ij}$ are voxel independent, the $j^\text{th}$ chemical reaction in a given voxel is of the same type as the $j^\text{th}$ chemical reaction in any other voxel. The diffusive interchange of chemicals between neighbouring voxels is modelled by the reactions:
\begin{equation}\label{RDMEdiffuse}
X_i^k \xrightleftharpoons[D_{i}]{D_{i}} X_i^{k'},~~k' \in Ne(k),
\end{equation}
where $D_i$ is the diffusion rate of species $X_i$ (which is independent of the voxel index and hence implies the assumption that the diffusion coefficient of a given species is space independent) and $Ne(k)$ is the set of voxels neighbouring $k$. The specific neighbours of a voxel depend on the topology of the lattice, though for the purposes of this paper it doesn't matter what this is, as long as every voxel is indirectly connected to every other by some path. The stochastic dynamics of this system is described by the RDME:
 \begin{align}\label{RDME}
\frac{d}{dt}P(\vec{n}^1,...,\vec{n}^M,t)=&\sum_{k=1}^M \sum_{j=1}^R  \left( \prod_{i=1}^N E_{i,k}^{s_{ij}-r_{ij}}-1\right)\hat{a}_j(\vec{n}^k,\frac{\Omega}{M})P(\vec{n}^1,...,\vec{n}^M,t) \nonumber\\
&+\sum_{k=1}^M\sum_{k' \in Ne(k)}\sum_{i=1}^N \left( E_{i,k}^{1}E_{i,k'}^{-1}-1\right)D_in_i^kP(\vec{n}^1,...,\vec{n}^M,t),
\end{align}
where $n_i^k$ is the number of molecules of species $X_i^k$, $\vec{n}^k=(n_1^k,...,n_N^k)^T$, $P(\vec{n}^1,...,\vec{n}^M,t)$ is the probability of the system being in state $ (\vec{n}^1,...,\vec{n}^M)$ at time $t$, and $E_{i,k}^x$ is the operator which replaces $n_i^k$ with $n_i^k+x$. Note that the first line of Eq. \eqref{RDME} refers to the chemical reactions while the second line refers to the diffusive reactions. The local propensity function is defined as: given that the state of voxel $k$ is $\vec{n}^k$, then $\hat{a}_j(\vec{n}^k,\frac{\Omega}{M}) dt$ is the probability that one reaction of index $j$ occurs somewhere inside this voxel in the next infinitesimal time interval $[t,t+dt)$.

\subsection{The RDME in the limit of fast diffusion}

We now investigate what happens to the RDME in the limit where $D_i \rightarrow \infty$ for all $i=1,...,N$. 
Suppose that the system is in state $(\vec{n}^1,...,\vec{n}^M)$ at time $t$, and define $n_i=n_i^1+...+n_i^M$ as the global molecule number of the species $X_i$. Each time a chemical reaction occurs somewhere in space, the local molecule number of one or more species changes leading to a corresponding change in the global number of molecules of the concerned species. Before the next chemical reaction occurs, diffusive reactions will happen a very large number of times such that the system will approach the steady-state of the purely diffusive system \eqref{RDMEdiffuse}. Specifically suppose that a chemical reaction just occurred somewhere in space and that the global state vector is $(n_1,n_2,...,n_N)$. Then it follows that due to the effect of infinitely fast diffusion, the probability of having $n_i^k$ molecules of species $X_i$ in voxel $k$, conditional on the fact that there are $n_i$ molecules of the same species in all of space, is given by the binomial distribution:
\begin{align}
\label{binfn}
P(n_i^k|n_i) = \frac{n_i!}{n_i^k! (n_i - n_i^k)!} \Bigl(\frac{1}{M}\Bigr)^{n_i^k} \Bigl(1 - \frac{1}{M}\Bigr)^{n_i - n_i^k}, \quad i = 1,...,N.
\end{align}   
It also follows that the distribution of molecule numbers of all chemical species in voxel $k$, conditional on the global state vector, is given by:
\begin{align}
\label{vdist}
P(\vec{n}^k|\vec{n}) = \prod_{i=1}^N P(n_i^k|n_i).
\end{align}
Note this distribution does take into account any implicit chemical conservation laws. This is since $P(\vec{n}^k)$ is conditional on the global state vector $(n_1,n_2,...,n_N)$ which is only changed when a chemical reaction occurs (since diffusion cannot cause a global change in the number of molecules but rather only induces a repartitioning of molecules across space). 

Now starting from the RDME, we want to calculate the probability $\tilde{a}_j(\vec{n},\Omega)dt$, in the limit of fast diffusion, that the $j^\text{th}$ chemical reaction occurs somewhere in the space of volume $\Omega$ in the next infinitesimal time interval $[t,t+dt)$, conditional on the global state vector, $\vec{n}=(n_1,n_2,...,n_N)$. This reaction can occur in either voxel 1 or voxel 2 or ... or voxel M and hence we must sum the propensities of the $j^\text{th}$ chemical reaction in each voxel. Furthermore we must also sum over all possible different states in each voxel which are compatible with the global state vector, $\vec{n}$. Hence it follows that:
\begin{align}
\label{micropropav1}
\tilde{a}_j(\vec{n},\Omega) = \sum_{k=1}^M \sum_{n_1^k=0}^{n_1} ... \sum_{n_N^k=0}^{n_N} \hat{a}_j(\vec{n}^k,\frac{\Omega}{M}) P(\vec{n}^k|\vec{n}).
\end{align}
If we use the simplifying assumption that the rate constant of the $j^{th}$ chemical reaction in a voxel is the same as the rate constant of the $j^{th}$ chemical reaction in any other voxel, then in the limit of fast diffusion, the average propensity of the $j^{th}$ chemical reaction in a given voxel is the same as that in any other voxel and hence Eq. (\ref{micropropav1}) further simplifies to:
\begin{align}
\label{micropropav2}
\tilde{a}_j(\vec{n},\Omega)&=M \sum_{n_1^k=0}^{n_1} ... \sum_{n_N^k=0}^{n_N} \hat{a}_j(\vec{n}^k,\frac{\Omega}{M}) P(\vec{n}^k|\vec{n}),
\end{align}
for some $k=1,...,M$. This can be written conveniently as an expected value under the Binomial distribution defined by Eq. \eqref{vdist}:
\begin{equation}
\tilde{a}_j(\vec{n},\Omega)=M\mathbb{E}\left[\hat{a}_j(\vec{n}^k,\frac{\Omega}{M})\right].
\end{equation}
 By the definition of the propensity function $\tilde{a}_j(\vec{n},\Omega)$, it follows immediately from the laws of probability that the master equation to which the RDME converges to, in the limit of fast diffusion, is:
\begin{equation} \label{EffectiveCME}
\frac{d}{dt}P(\vec{n},t)=\sum_{j=1}^R  \left(\prod_{i=1}^N E_i^{s_{ij}-r_{ij}}-1\right)\tilde{a}_j(\vec{n},\Omega)P(\vec{n},t).
\end{equation}
Note that this equation is identical to Eq. \eqref{CME} except that $\hat{a}_j$ is replaced with $\tilde{a}_j$. The main result follows: \emph{The RDME converges to the CME in the limit of fast diffusion if $\tilde{a}_j(\vec{n},\Omega)=\hat{a}_j(\vec{n},\Omega)$} for all $j=1,...,R$.

With this in mind, we can define a class of propensities which we call \emph{convergent propensities}. A propensity function $\hat{a}_j(\vec{n},\Omega)$ is convergent if $\tilde{a}_j(\vec{n},\Omega)=\hat{a}_j(\vec{n},\Omega)$. A system consisting exclusively of convergent propensities will have the satisfying property of convergence of the RDME to the CME, in the fast diffusion limit. A system with at least one non-convergent propensity will most likely not have this property, though this cannot be generally excluded. 

\subsection{The RDME and the CME in the limits of fast diffusion and large volumes}\label{LNAc}

We now briefly show that a non-convergent RDME can still converge to the correct CME in the limit of fast diffusion, if we furthermore take the macroscopic limit of large volumes. The proof is based on the fact that in the macroscopic limit, the solution of a master equation is approximated by the chemical Langevin equation whose solution has sharp peaks centred on the solution of the corresponding deterministic rate equations (REs) \cite{Gillespie2009}. 

The REs of system \eqref{CMEreact} described stochastically by the CME are given by the equation:
\begin{equation}\label{CMERE}
\frac{d}{dt} \vec{\phi}=\sum_{j=1}^R (r_{ij}-s_{ij})f_j(\vec{\phi}),
\end{equation}
where $\vec{\phi} = \langle \vec{n} \rangle /\Omega$ is the vector of deterministic concentrations of species $X_1,...,X_N$ (the angled brackets signify the average taken in the macroscopic limit), and $f_j(\vec{\phi})$ is the \emph{macroscopic propensity function} of reaction $j$ defined as:
\begin{equation}\label{fdef}
f_j(\vec{\phi})=\text{lim}_{\Omega \rightarrow \infty}\frac{\hat{a}_j(\vec{n}=\Omega \vec{\phi},\Omega)}{\Omega}.
\end{equation}
The REs of the system given by Eqs. \eqref{RDMEreact} and \eqref{RDMEdiffuse}, described stochastically by the RDME, are given by:
\begin{equation}\label{RDMERE1}
\frac{d}{dt}\phi_i^k=\sum_{j=1}^R (r_{ij}-s_{ij})f_j(\vec{\phi}^k) + D_i \biggl(\sum_{k'} \phi_i^{k'} - z \phi_i^k \biggr),
\end{equation}
where $\vec{\phi}^k = \langle \vec{n}^k \rangle /(\Omega/M)$ is the vector of deterministic concentrations in voxel $k$ and $f_j$ is the same $f_j$ as defined in Eq. \eqref{fdef}. Note that the sum over $k'$ is a sum over the set of voxels neighbouring $k$, and $z$ is the number of neighbours of a voxel for a particular RDME lattice. Clearly in the limit of fast diffusion, the deterministic concentration of each species is the same in all voxels, i.e., $\phi_i^k=\phi_i^{k'}$ at all times. Hence the second term on the right hand side of Eq. (\ref{RDMERE1}) equals zero and the rate equations of the RDME simplify to:
\begin{equation}\label{RDMERE}
\frac{d}{dt}\phi_i^k=\sum_{j=1}^R (r_{ij}-s_{ij})f_j(\vec{\phi}^k).
\end{equation}
Now to compare with the REs of the CME, we need to use Eq. (\ref{RDMERE}) derive the REs for the global concentration $\phi_i$, which follows from the definition of the global molecule numbers and is given by $\phi_i=\frac{1}{M}(\phi_i^1+...+\phi_i^M)$. These are given by:
\begin{equation}\label{RDMEREGLOB}
\frac{d}{dt}\phi_i=\frac{1}{M}\frac{d}{dt} \sum_{k=1}^M \phi_i^k=\frac{1}{M}\sum_{k=1}^M \sum_{j=1}^R(r_{ij}-s_{ij})f_j(\vec{\phi}^k)=\sum_{j=1}^R (r_{ij}-s_{ij})f_j(\vec{\phi}),
\end{equation}
where the last equation follows from the fact that in the fast diffusion limit, $\phi_i^k=\phi_i$ at all times. The RE of the global concentrations of the RDME given by Eq. \eqref{RDMEREGLOB} is therefore equal to Eq. \eqref{CMERE}, the RE of the global concentrations of the CME. In summary, \emph{in the combined limits of fast diffusion and large volumes, the RDME of a system converges to the CME of the same system, regardless of whether the propensities are convergent.}

It can also be straightforwardly shown using the linear noise approximation (LNA) that for deterministically monostable systems, the solution of the RDME and CME, in the macroscopic and fast diffusion limits, tends to the same Gaussian centred on the RE solution \cite{VKbook}. This follows from the fact that the variance and covariance of the Gaussian are functions of the RE solution and which, as we have shown, is one and the same for the RDME and CME.

\section{Examples of convergent and non-convergent propensities}\label{Converg}

In this section we test whether some commonly-used propensities are in the convergent or non-convergent class. We shall assume, for simplicity, that the rate constant of the $j^{th}$ chemical reaction in a voxel is the same as the rate constant of the $j^{th}$ chemical reaction in any other voxel. 

\subsection{Convergent propensities}

Consider mass-action kinetics. It then follows \cite{VKbook} that the propensity of the $j^{th}$ chemical reaction in the CME and of the $j^{th}$ chemical reaction in voxel $k$ of the RDME are respectively given by:
\begin{align}
\hat{a}_j(\vec{n},\Omega)&=\Omega k_j \prod_{z=1}^N \Omega^{-s_{zj}} \frac{n_z!}{(n_z - s_{zj})!}, \\ \label{microp1}
\hat{a}_j(\vec{n}^k,\frac{\Omega}{M})&=\frac{\Omega}{M} k_j \prod_{z=1}^N \biggl(\frac{\Omega}{M}\biggr)^{-s_{zj}}\frac{n_z^k!}{(n_z^k - s_{zj})!}.
\end{align} 
The most commonly used types of reactions following mass-action kinetics are the zero order reaction, first order reaction, second-order reactions between similar reactants and second-order reaction with different reactants which have CME propensities $\hat{a}_j(\vec{n},\Omega)$ equal to $k_j \Omega$, $k_j n_i$, $(k_j /\Omega) n_i (n_i-1)$ and $(k_j /\Omega) n_i n_j$ respectively, for some integers $i$ and $j$.

Substituting Eq. (\ref{microp1}) in Eq. (\ref{micropropav2}), we have that in the fast diffusion limit, the RDME converges to a master equation with a propensity for the $j^{th}$ chemical reaction equal to:
\begin{align}
\tilde{a}_j(\vec{n},\Omega)&=M \prod_{z=1}^N  \sum_{n_z^k=0}^{n_z} \hat{a}_j(\vec{n}^k,\frac{\Omega}{M}) P(n_z^k|n_z), \\ \label{effpropma}
&=M \frac{\Omega}{M} k_j \prod_{z=1}^N \biggl(\frac{\Omega}{M}\biggr)^{-s_{zj}} \sum_{n_z^k=0}^{n_z} \frac{n_z^k!}{(n_z^k - s_{zj})!} P(n_z^k|n_z).
\end{align}
Now the quantity:
\begin{align}
 \sum_{n_z^k=0}^{n_z} \frac{n_z^k!}{(n_z^k - s_{zj})!} P(n_z^k|n_z),
\end{align}
is by definition the $s_{zj}$-th factorial moment of the binomial distribution $P(n_z^k|n_z)$ with success probability $1/M$ and number of trials $n_z$. This factorial moment is a standard result (see for example \cite{Potts1953}) and is given by:
\begin{align}
\sum_{n_z^k=0}^{n_z} \frac{n_z^k!}{(n_z^k - s_{zj})!} P(n_z^k|n_z) = \frac{n_z!}{(n_z - s_{zj})!} \biggl(\frac{1}{M}\biggr)^{s_{zj}}.
\end{align}
Substituting the above equation in Eq. (\ref{effpropma}) we obtain:
\begin{align}
\tilde{a}_j(\vec{n},\Omega)=\Omega k_j \prod_{z=1}^N \Omega^{-s_{zj}} \frac{n_z!}{(n_z - s_{zj})!}=\hat{a}_j(\vec{n},\Omega).
\end{align}
It follows that the propensities of reactions following mass-action are convergent; this applies to all reaction orders. The convergence of the RDME to the CME in the fast diffusion limit, for reactions up to second-order, has been previously also shown by Gardiner \cite{gardiner1985handbook} using a completely different method.

\subsection{Non-convergent propensities}
It is common practice to use effective propensities which lump a number of elementary reactions together. One of the most popular of such propensities is the Michaelis-Menten propensity. This can model various processes such as nonlinear degradation of a protein, enzyme catalysis of a protein into a product or the activation of a gene by a protein. Let this protein species be $X_i$. If the $j^{th}$ reaction is of the Michaelis-Menten type, then it can be described by a term in the deterministic rate equations of the form $f_j(\vec{\phi}) = \frac{k_j \phi_i}{K+\phi_i}$. Using Eq. (\ref{fdef}), one can deduce that a corresponding effective propensity in the CME would be $\hat{a}_j(\vec{n},\Omega)=\frac{k_j n_i}{K+n_i/\Omega}$ \cite{Gonze2002,thomas2012a}. The corresponding propensity in the $k^{th}$ voxel of the RDME would be $\hat{a}_j(\vec{n}^k,\frac{\Omega}{M})=\frac{k_j n_i^k}{K+M n_i^k/\Omega}$ \cite{Lawson2015}. 

Substituting the latter propensity of the RDME in Eq. (\ref{micropropav2}), we have that in the fast diffusion limit, the RDME converges to a master equation with a propensity for the $j^{th}$ chemical reaction equal to:
\begin{align}
\tilde{a}_j(\vec{n},\Omega)&=M k_j \sum_{n_i^k=0}^{n_i} \frac{n_i^k}{K+M n_i^k/\Omega} P(n_i^k|n_i), \nonumber \\ \label{MMeffpropn}
&=\frac{k_j\Omega(M-1)^{n_i-1}n_i~_2F_1\left(K\Omega/M+1,1-n_i;K\Omega/M+2;-1/(M-1)\right)}{M^{n_i}(K\Omega/M+1)}\neq \frac{k_jn_i}{K+n_i/\Omega} = \hat{a}_j(\vec{n},\Omega),
\end{align}
where $~_2F_1(a,b;c;d)$ is a hypergeometric function.
Since $\tilde{a}_j(\vec{n},\Omega) \neq  \hat{a}_j(\vec{n},\Omega)$, it follows that Michaelis-Menten propensities are non-convergent. However note in the limits of small $n_i^k$ and of very large $n_i^k$, $\hat{a}_j(\vec{n}^k,\frac{\Omega}{M})$ reduces to $k_j n_i^k / K$ and $k_j \Omega / M$ respectively; these are special cases of mass-action kinetics Eq. (\ref{microp1}) and hence in these limits, the Michaelis-Menten propensity can be considered convergent. The largest deviations from mass-action occur when $n_i^k/\Omega$ is roughly equal to the constant $K$ (the Michaelis-Menten constant); this is the case where the non-convergence of the Michaelis-Menten propensity becomes most apparent. 

A generalisation of the Michaelis-Menten propensity, which is sometimes used, is given by the Hill propensity. For the CME this is given by $\hat{a}_j(\vec{n},\Omega)=\frac{\Omega k_j(n_i)^\theta}{(\Omega K)^\theta+(n_i)^\theta}$, for some $i$. The Hill coefficient $\theta$ is  an integer greater or equal to 1. The corresponding propensity in voxel $k$ of the RDME is $\hat{a}_j(\vec{n}^k,\frac{\Omega}{M})=\frac{(\Omega / M) k_j(n_i^k)^\theta}{(\Omega K / M)^\theta+(n_i^k)^\theta}$. For general $\theta$ it is not possible to evaluate Eq. (\ref{micropropav2}) analytically. However, if we can find a set of parameters for which $\tilde{a}_j$ and $\hat{a}_j$ do not agree, then we can be certain that they are not the same function. Choose, for instance, $k_1=K=\Omega=1$, $M=n_i=2$. For these parameters, $\hat{a}_j=\frac{2^\theta}{1+2^{\theta}}$. On the other hand, $\tilde{a}_j=\frac{1}{2} \left( \frac{1}{2}+\frac{1}{(1/2)^\theta+1}+\frac{1}{2}\frac{2^\theta}{(1/2)^\theta+2^\theta}\right)$. There are no real values of $\theta$ for which $\hat{a}_j=\tilde{a}_j$, and therefore it follows that Hill-type propensities are non-convergent.

Another type of propensity related to the ones described above is that used to effectively model the repression of a gene by a protein $X_i$. In the CME, this propensity is given by $\hat{a}_j(\vec{n},\Omega)=k_j \Omega \frac{(K \Omega)^{\theta}}{(K \Omega)^{\theta}+n_i^{\theta}}$ \cite{Gonze2002}. Writing down the RDME equivalent of this propensity and using Eq. (\ref{micropropav2}), one can also show that this propensity is non-convergent. 


\section{Simple convergent and non-convergent example systems}\label{Examples}

To illustrate the results of this paper, we briefly apply them to some simple example systems in this section. For simplicity we will use systems consisting of only one species.

\subsection{A convergent example}

A simple convergent example is given by the open dimerisation reaction:
\begin{equation}\label{Dimer}
\emptyset \rightarrow X,~X+X \rightarrow \emptyset.
\end{equation}
The propensity functions of the CME are:
\begin{equation}
\hat{a}_1(n,\Omega)=k_1\Omega,~\hat{a}_2(n,\Omega)=\frac{k_2}{\Omega}n(n-1),
\end{equation}
where $n$ is the number of $X$ molecules. The corresponding propensity functions in voxel $k$ of the RDME are obtained by replacing $n$ by $n^k$ and $\Omega$ by $\Omega/M$. 	

\begin{figure}[h]
\includegraphics[scale=0.3]{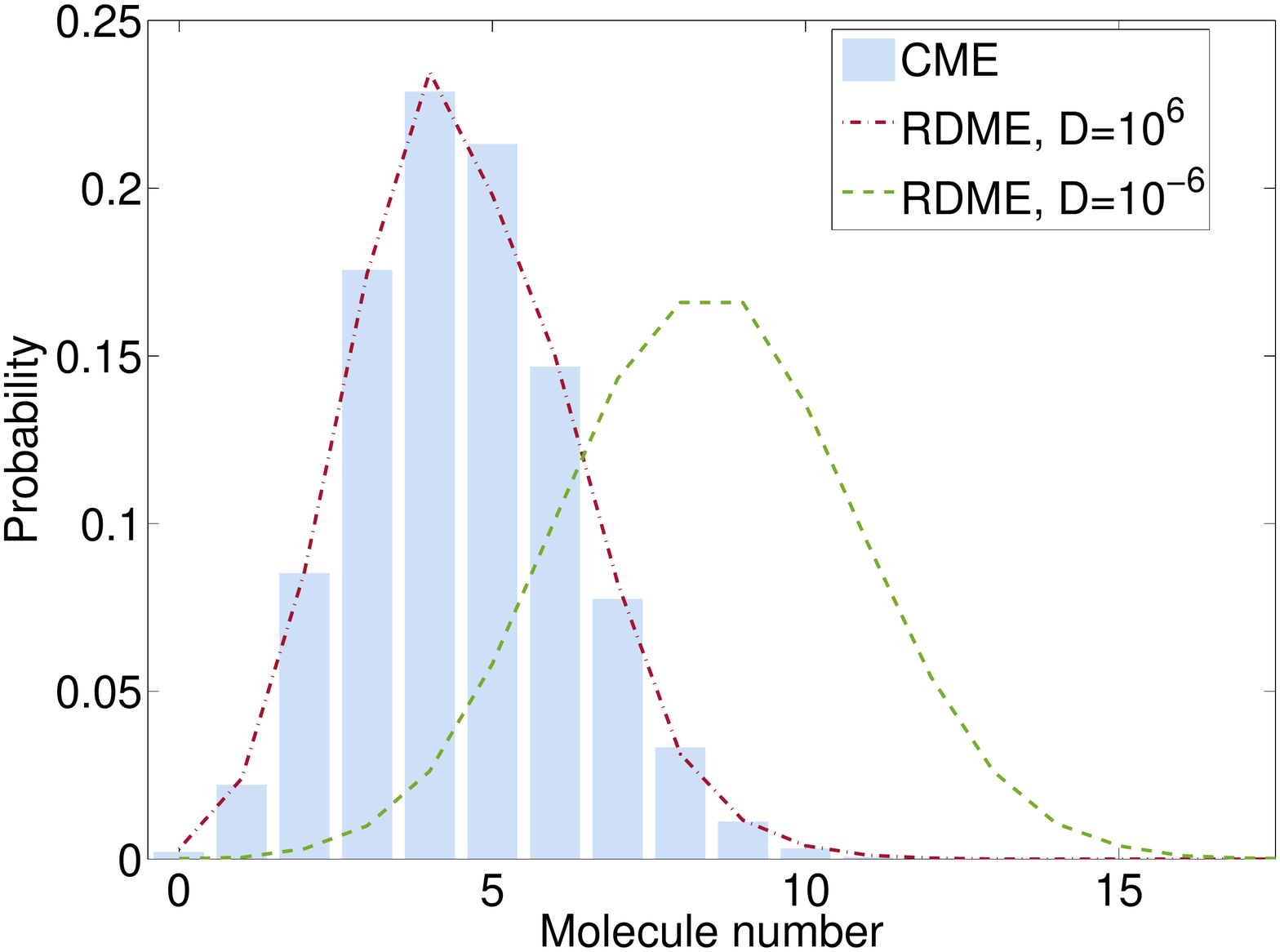}
\caption{Steady state solution of the CME for system \eqref{Dimer} (blue histogram) compared with the steady-state solutions of the global molecule numbers calculated using the RDME for very slow diffusion (green dashed line) and for very fast diffusion (red dash-dotted line), all obtained with the SSA. Note that the RDME converges to the CME in the limit of fast diffusion, as predicted by our theory. Parameter values are $k_1=1000$, $k_2=30$, $\Omega=1$. The number of voxels of the RDME is $M=10$.}\label{fig1}
\end{figure}

By the results of Section \ref{Converg}A, since both propensities are of the mass-action type, they are convergent. In Fig. \ref{fig1} we show that the the steady-state distribution of global molecule numbers (sum of molecule number over all voxels) calculated using the RDME with very slow diffusion ($D = 10^{-6}$) disagrees with the CME while the same computed with very fast diffusion ($D = 10^{6}$) agrees exactly with the CME. The plots are obtained using the stochastic simulation algorithm (SSA) \cite{gillespie1977exact}. The agreement of the CME and RDME in the limit of fast diffusion is in agreement with the theoretical result that both propensities are convergent. 

\subsection{A non-convergent example}

A simple non-convergent example is given by a protein production and nonlinear protein degradation system:
\begin{equation}\label{MM}
\emptyset \rightarrow X,~X \rightarrow \emptyset.
\end{equation}
The propensity functions in the CME are:
\begin{equation}
\hat{a}_1(n,\Omega)=k_1\Omega,~\hat{a}_2(n,\Omega)=\frac{k_2n}{K+n/\Omega}.
\end{equation}
The corresponding propensity functions in voxel $k$ of the RDME are obtained by replacing $n$ by $n^k$ and $\Omega$ by $\Omega/M$. 

By the results of Section \ref{Converg}A and B, we see that the first propensity is convergent but the second is not. It follows that the RDME will not converge to the CME in the fast diffusion limit. In particular, using the method derived in Ref. \cite{smith2015general} for general one variable, one step-processes,  one can show that the exact steady-state analytical solutions of the CME (Eq. (\ref{CME}) with the above propensities) and of the master equation to which the RDME converges in the fast diffusion limit (Eq. (\ref{EffectiveCME}) with $\tilde{a}_1(n,\Omega)=k_1\Omega$ and $\tilde{a}_2(n,\Omega) $ given by Eq. (\ref{MMeffpropn})) are given by:
\begin{equation}\label{exact2}
P(0)=C_1,~P(n)=C_1\prod_{i=1}^n \frac{k_1\Omega(K+i/\Omega)}{k_2i},
\end{equation}
and
\begin{equation}\label{exact}
P(0)=C_2,~P(n)=C_2\prod_{i=1}^n \frac{k_1 M^{i}(K\Omega/M+1)}{k_2(M-1)^{i-1}i~_2F_1\left(K\Omega/M+1,1-i;K\Omega/M+2;-1/(M-1)\right)},
\end{equation}
respectively, where $C_1$ and $C_2$ are normalisation constants.

\begin{figure}[h]
\centering
\includegraphics[scale=0.3]{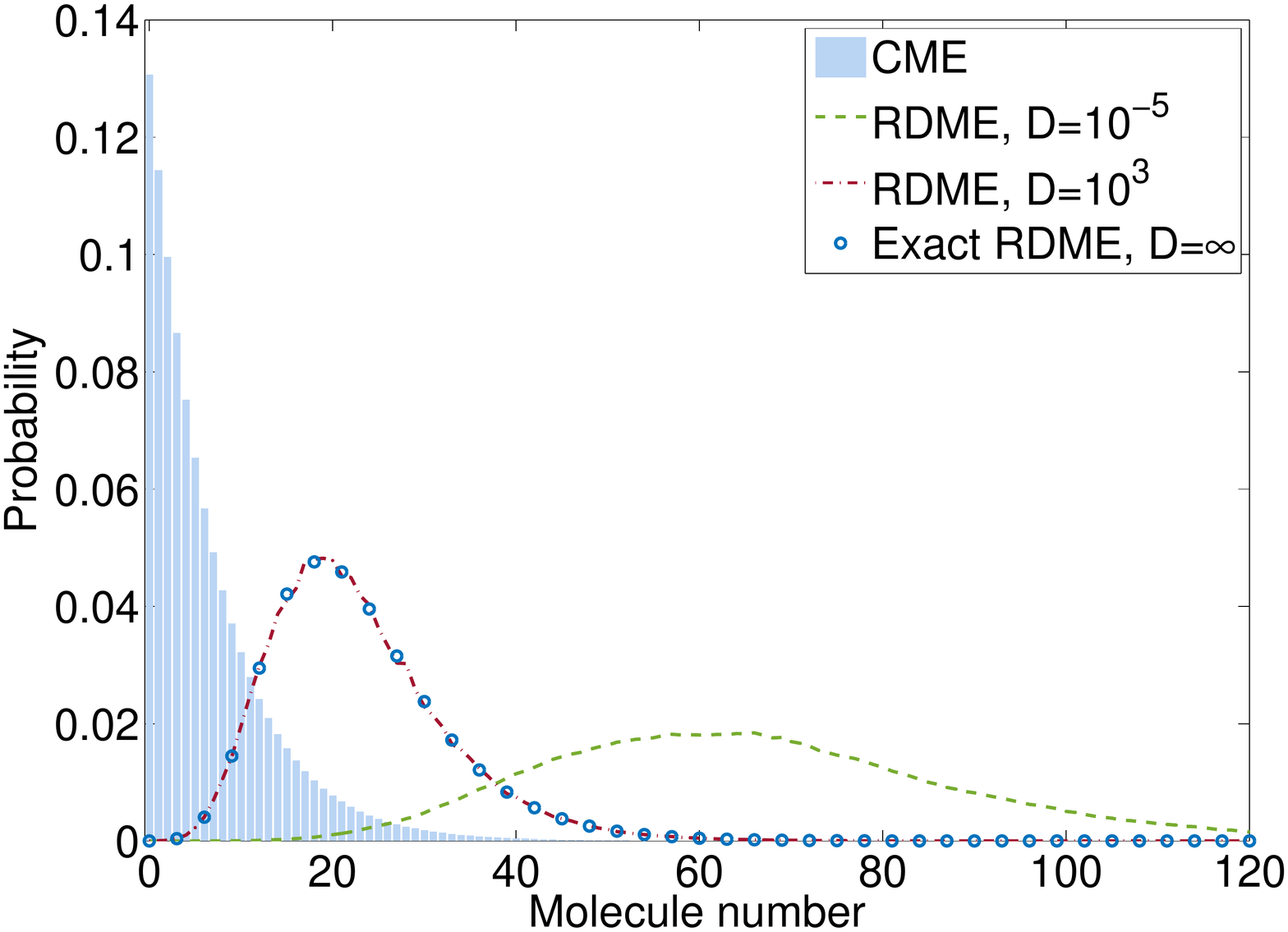} \includegraphics[scale=0.3]{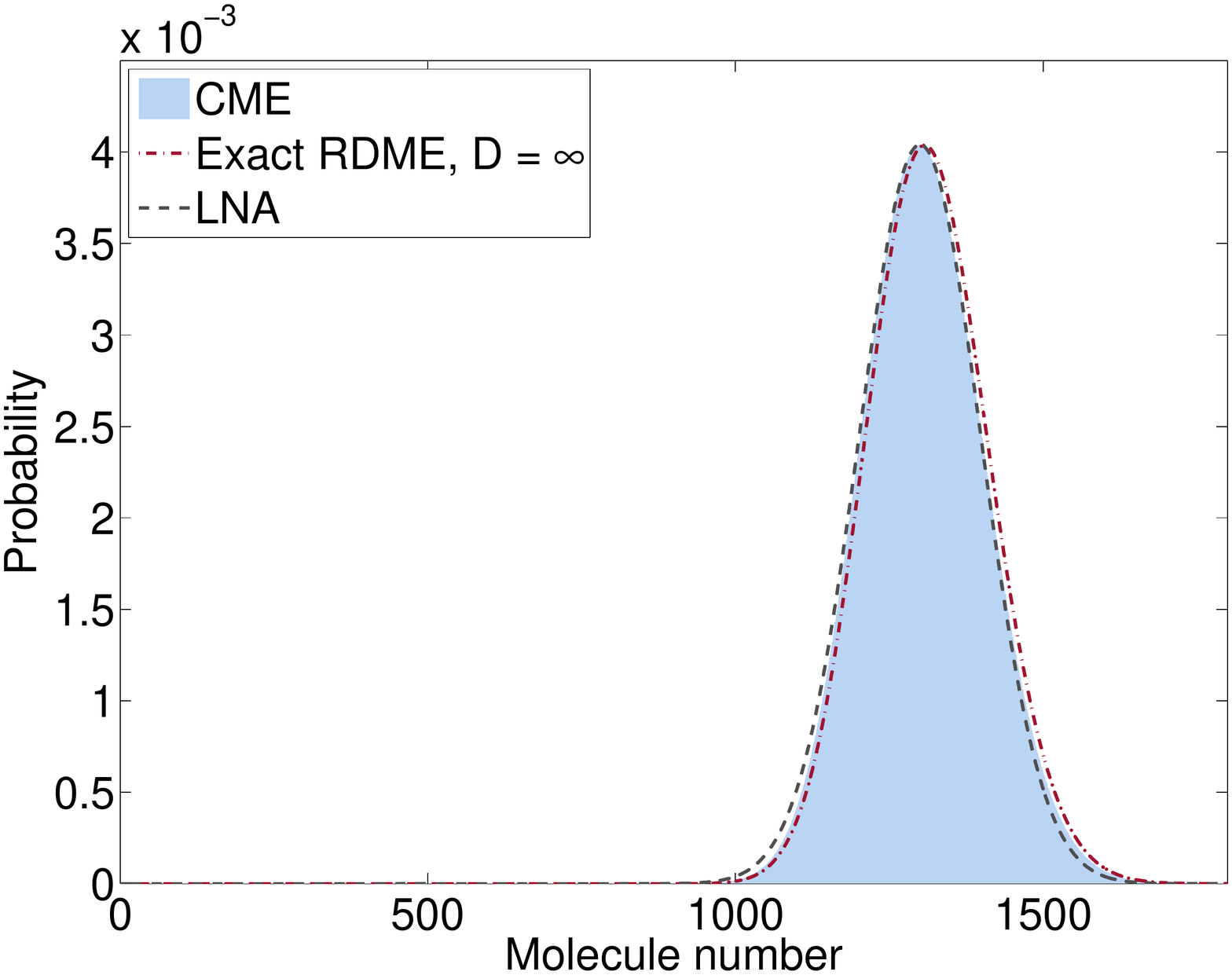}\\
(a)~~~~~~~~~~~~~~~~~~~~~~~~~~~~~~~~~~~~~~~~~~~~~~~~~~~~~~~~~~~~~~~~~~~~~~~~~~~~~~~~(b)
\caption{(a) Steady state solution of the CME for system \eqref{MM} (blue histogram) compared with the RDME solutions for slow diffusion (green dashed line) and for fast diffusion (red dash-dotted line), all obtained with the SSA, and the exact RDME solution for infinite diffusion given by Eq. \eqref{exact} (blue circles). (b) Steady state exact solution of the CME given by Eq. \eqref{exact2} compared with the steady state exact solution of the RDME for infinite diffusion given by Eq. \eqref{exact} and the LNA for system \eqref{MM}. Parameter values are $k_1=2.6$, $k_2=3$, $K=0.01$. The number of voxels in the RDME is $M=10$. The total volume is $\Omega=1$ in (a) and $\Omega=20000$ in (b).}\label{fig2}
\end{figure}

In Fig. \ref{fig2}(a) we verify using the SSA that for a small volume $\Omega=1$, the RDME disagrees with the CME for both very slow and very fast diffusion, i.e., the RDME does not converge to the CME in limit of fast diffusion. Note also that the exact solution of the master equation to which the RDME converges in the fast diffusion limit as given by Eq. (\ref{exact}) agrees very well with that obtained using stochastic simulations of the RDME for large diffusion $D=10^3$; this agreement is an independent verification of our theory. 

In Fig. \ref{fig2}(b) we show that for large volumes $\Omega = 20000$, the stochastic simulations agree very well with the exact RDME solution Eq. (\ref{exact}) at $D=\infty$ and with the exact CME solution Eq. (\ref{exact2}). In this case the RDME and CME are the Gaussian distribution centred on the solution of the rate equations which is predicted by the LNA. This is in agreement with the results of Section \ref{LNAc} and shows that the non-convergence problems of the RDME are mostly relevant at small volumes (or equivalently small molecule numbers). 

\section{Discussion}\label{Discussion}
In this paper we have proved a remarkable and counter-intuitive fact, namely that the RDME does not necessarily converge to the CME in the limit of fast diffusion. This has serious consequences for the validity of the RDME in general, since it cannot accurately predict behaviour that should be inside its area of applicability, namely systems with fast diffusion and non-elementary reactions. 

Consequences of this fact for mathematical modelling of biology are numerous. The class of non-convergent propensities includes Michaelis-Menten-type rates, which are frequently used to model metabolic systems, and Hill-type rates, which describe transcription factor binding. The results of this paper suggest that spatial models of genetic or metabolic networks with the RDME are likely to give both quantitatively and qualitatively incorrect results. In particular though the differences between the RDME in the fast diffusion and CME disappear in the limit of large volumes, this limit is not typically relevant to the modelling of intracellular kinetics because of the sheer small cellular volume \cite{thomas2013reliable}. 

Future work in this area may aim at fixing the non-convergence problem. A particularly interesting idea would be to derive a set of special propensity functions for the RDME which converge to the CME propensities in the limit of fast diffusion. This would ensure a consistent way of performing spatial and stochastic simulations. 

\bibliography{mybibfile}{}
\end{document}